\documentclass[%
reprint,
 amsmath,amssymb,
 aps,
pra,
showkeys
]{revtex4-1}

\usepackage{graphicx}
\usepackage{dcolumn}
\usepackage{bm}

\usepackage{setspace}

\usepackage{amsthm}
\usepackage{mathrsfs}
\usepackage{color}
\usepackage[hyperindex,breaklinks,bookmarks=true,pdfstartview={FitH}]{hyperref}
\usepackage[figure]{hypcap}

\usepackage{xcolor}

\usepackage{xfrac}
\usepackage{dsfont}
\usepackage{bbm}

\newcommand{\tr}[1]{\mathrm{Tr}\left[#1\right]}
\newcommand{\ket}[1]{\vert #1 \rangle}
\newcommand{\bra}[1]{\left\langle #1 \right\vert}

\newcommand{\ip}[2]{\langle #1|#2\rangle}         
\newcommand{\vect}[1]{\mathbf{#1}}     

\newcommand{\be}{\begin{equation}}
\newcommand{\eq}{\end{equation}}

\usepackage[utf8]{inputenc}

\DeclareFontFamily{U}{mathb}{\hyphenchar\font45}
\DeclareFontShape{U}{mathb}{m}{n}{
<-6> mathb5 <6-7> mathb6 <7-8> mathb7
<8-9> mathb8 <9-10> mathb9
<10-12> mathb10 <12-> mathb12
}{}
\DeclareSymbolFont{mathb}{U}{mathb}{m}{n}
\DeclareMathSymbol{\succt}{\mathrel}{mathb}{"CF}

\newcommand{\AC}{\mathcal{A}}
\newcommand{\BC}{\mathcal{B}}

\newcommand{\DC}{\mathcal{D}}

\newcommand{\JC}{\mathcal{J}}

\newcommand{\SC}{\mathcal{S}}

\newcommand{\UC}{\mathcal{U}}

\newcommand{\ii}{\mathrm{i}}					  
\newcommand{\expo}[1]{\mathrm{e}^{#1}} 
\newcommand{\expoi}[1]{\expo{\ii #1}} 

\begin{document}

\title{Uncertainty, joint uncertainty, and the quantum uncertainty principle} 

\author{Varun Narasimhachar}
\email{vnarasim@ucalgary.ca}
\author{Alireza Poostindouz}
\email{alireza.poostindouz@ucalgary.ca}
\author{Gilad Gour}
\email{gour@ucalgary.ca}
\affiliation{Department of Mathematics and Statistics and Institute for Quantum Science and Technology, University of Calgary, 2500 University Drive NW, Calgary, Alberta, Canada T2N 1N4}

\date{March 10, 2016}

\begin{abstract}
Historically, the element of uncertainty in quantum mechanics has been expressed through mathematical identities called uncertainty relations, a great many of which continue to be discovered. These relations use diverse measures to quantify uncertainty (and joint uncertainty). In this paper we use operational information-theoretic principles to identify the common essence of all such measures, thereby defining measure-independent notions of uncertainty and joint uncertainty. We find that most existing entropic uncertainty relations use measures of joint uncertainty that yield themselves to a small class of operational interpretations. Our notion relaxes this restriction, revealing previously unexplored joint uncertainty measures. To illustrate the utility of our formalism, we derive an uncertainty relation based on one such new measure. We also use our formalism to gain insight into the conditions under which measure-independent uncertainty relations can be found.
\end{abstract}

\maketitle 

\section*{Introduction}

Revealing one of the most striking features of quantum mechanics, Heisenberg \cite{Heisenberg} showed that the outcomes of certain pairs of measurements on a quantum system can never be predicted simultaneously with certainty{\textemdash}regardless of how the system is prepared. Heisenberg's original statement of what he called the ``indeterminacy'' principle concerned potential measurements of the position and the momentum of a quantum particle. Many later works \cite{K1927,Robertson1929,Hirschman1957,Beckner1975,Biaynicki-Birula1975} lent quantitative rigor to Heisenberg's original idea and generalized it, both in the number and type of measurements involved and in the measures used to quantify joint uncertainty. At the same time, Heisenberg himself set off another chain of research on a related concept: measurement-induced disturbance and so-called noise-disturbance relations \cite{Heis2,ErDis1,ErDis2,ErDis3}.

Pioneered by Hirschman \cite{Hirschman1957}, many works \cite{Deutsch1983,Maassen1988,Ivanovic1992,Sanchez1993,Ballester2007,Wu2009,huang1,huang2,note11} have used entropies to quantify uncertainty, culminating in a recent surge of quantum information-theoretic treatments of the uncertainty principle \cite{Berta2010,Tomamichel2011,coles11,Coles2012,Friedland2013,Puchaa2013,DSum,DSC,coles2014,bcw14,KTW14,qcur}. An important contribution of these recent works is the formulation of uncertainty relations applicable on a quantum system correlated with a quantum memory; such relations are used to strengthen the security proofs of cryptographic tasks \cite{QKD1,QKD2}. These are in addition to existing applications of uncertainty relations in quantum cryptography \cite{DiVincenzo2004,Damgaard2005,Koashi2005}, the study of quantum nonlocality \cite{Oppenheim2010,Guhne2004}, and continuous-variable quantum information processing \cite{Squeezed,CVQIP1,CVQIP2,cv12}.

In all of these areas, the primary ingredient is the concept of the uncertainty of a variable, as well as that of the joint uncertainty of several variables. The aim of this paper is to clarify these concepts from an information-theoretic perspective. In the literature, the uncertainty of a variable has almost always been discussed in terms of measures that quantify ``the amount of uncertainty'', e.g. the Shannon entropy and its extended family of R\'enyi entropies, geometric norm-based measures such as the quadratic variance, etc. In most cases, there is a clear operational meaning for such measures, rendering them well-suited to the particular application wherein they are used. Similarly, measures of the joint uncertainty of more than one variable have been constructed either by considering operational tasks that involve all the variables, or by combining single-variable uncertainty measures mathematically. In the present work we extract the common thread beneath the operational descriptions of all such (single or joint) uncertainty measures, resulting in some basic operational axioms that are independent of the measure used to quantify uncertainty, and that define the essence of our concept of uncertainty.

These axioms are motivated by information-theoretic principles that are intended to be as objective as possible. Considering the challenges inherent in such a requirement, we restrict the generality of our treatment in the following ways. Firstly, we restrict to notions of uncertainty applied to classical random variables. In particular, this class of variables includes the classical outcomes of quantum-mechanical measurements. Secondly, we avoid measures of uncertainty that explicitly involve the values of a variable, and instead consider only such measures that depend on the variable's \emph{probability distribution}. This necessitates a restriction to discrete variables; in fact, we consider only finite-dimensional variables. We make some tentative suggestions for the treatment of discrete and continuous infinite-dimensional cases, but leave the actual extension for future work. Finally, in comparing the uncertainties of different variables (which a measure of uncertainty should naturally be expected to enable), we will require the compared variables to represent the same type of physical quantity. For example, a comparison between the uncertainties in two different length variables will be possible within our formalism, but not one between a length uncertainty and a mass uncertainty.

The crux of this paper are the following axioms: (1) One's knowledge about a variable cannot increase under any processing without addition of new information about the variable; (2) The uncertainty in a variable representing a physical observable is invariant under the symmetries of the observable; and (3) The joint uncertainty of several variables is a valid concept even without an underlying operational description that combines those variables.

The first two axioms are inspired by earlier approaches \cite{Friedland2013,Puchaa2013,DSum,DSC} to measure-independent notions of uncertainty, wherein the connection between uncertainty and a mathematical concept called majorization \cite{Marshall2011} was utilized. Majorization is a hierarchy among probability distributions, induced by the action of a class of transformations called doubly stochastic maps. In this paper, by finding a mathematical characterization of mechanisms that can increase a variable's uncertainty, we gain an operational understanding of why, and to what extent, majorization plays a role in characterizing uncertainty.

First, we find that for variables with unrestricted symmetries, uncertainty-increasing mechanisms are associated with the set of \emph{all} doubly stochastic maps, leading to the emergence of majorization as the relation determining uncertainty. A function that quantifies uncertainty must then possess the property of never decreasing under any doubly stochastic maps. On the other hand, with restricted symmetries, only certain sub-classes of doubly stochastic matrices feature. The resulting hierarchy is then different from majorization, and a measure of uncertainty is required to be non-decreasing only under the restricted classes of doubly stochastic maps. This opens up more options for functions that can serve as uncertainty measures for variables with restricted symmetries.

Another element of novelty in our work lies in the third of our axioms, concerning joint uncertainty. In the context of physics, we can rephrase this axiom in terms of experiments: Suppose that we are interested in quantifying the joint uncertainty of several experiments, e.g. in connection with the quantum uncertainty principle, where the several experiments are different quantum measurements. One approach would be to construct new experiments that combine the original experiments in some way. For example, consider the following combined experiments constructed from a given set of experiments: (a) all the original experiments are performed independently; (b) all the apparatuses are set up, but only one of the experiments is chosen at random and performed.

The uncertainty in the outcome of such a combined experiment would quantify the joint uncertainty of the constituent experiments. But we see that there are different ways to combine experiments, which all capture different aspects of the joint uncertainty. In this paper we argue that the richness of joint uncertainty is not captured even by considering all such combined experiments. The most general notion of joint uncertainty is devoid of the particulars of such combinations, and allows all the component experiments to be, in principle, counterfactual. To illustrate this, we consider an extensively-studied type of quantum uncertainty relations: the so-called preparational uncertainty relations. For ease of explanation, let's consider a two-measurement preparational uncertainty relation, which has the generic form
\be\label{iup}
\JC\left(\vect p(\rho),\vect q(\rho)\right)\ge c,
\eq
where $\JC$ is a measure of the joint uncertainty of two variables, and $\vect p(\rho)$ and $\vect q(\rho)$ are the expected outcome probability distributions of a pair of measurements performed on a quantum state represented by the density operator $\rho$ (our arguments can be extended to more than two measurements). We show that most existing preparational uncertainty relations can be subjected to one of the specific operational interpretations (a) and (b) mentioned above. To show that these two interpretations are unnecessarily restrictive, we construct joint uncertainty measures that cannot be interpreted either way. We go on to derive an uncertainty relation based on one such measure, which is a relation nontrivially different from all the ones discovered in the past. The main purpose of deriving this new relation is to demonstrate the possibilities opened up by our joint uncertainty axiom.

Another contribution of this paper is a deeper understanding of so-called \emph{universal uncertainty relations} found in~\cite{Friedland2013,Puchaa2013,DSum,DSC}: pairs of vectors $(\vect u,\vect v)$ such that $\JC(\vect u,\vect v)$ provides a nontrivial bound [like the $c$ of Eq.~(\ref{iup})] for a whole class of measures, $\JC\in\mathsf J$. We find that no universal relations exist if $\mathsf J$ includes all possible measures; however, restricting to specific operational frameworks [using the (1) and (2) types of combined experiments discussed in the previous paragraph] is what makes the nontrivial universal relations found in \cite{Friedland2013,Puchaa2013,DSum,DSC} possible.

Even though we focus on preparational uncertainty relations in quantum mechanics, in principle our notions can be applied to any situation where probability-based uncertainty measures of classical variables are relevant. We summarize the possible applications in the conclusion, along with open problems.

\section{What is uncertainty?}\label{Sec:Uncertainty}
We will now develop a notion of uncertainty that can be applied to finite-dimensional classical variables. In particular, we seek a general method of comparing the uncertainties of two variables, in such a way that the comparison gives the same verdict independent of the function used to measure uncertainty. For our purpose, it will be sufficient to be able to compare physically similar variables, that is, variables representing the same underlying physical quantity; for example, comparing the uncertainty of a length with that of another length. We will not concern ourselves with how a comparison can be made between dissimilar variables.

Consider an experiment where Alice is about to roll a (possibly biased) die, whose faces she calls ``1'', ``2''\dots, ``$6$''. The eventual outcome of the roll will be a value $x\in\{1,2\dots,6\}$, but since we don't know $x$ a priori, we represent it as a random variable $X\equiv\{(x,p_x)\}$. What is \emph{Alice's minimum uncertainty about $X$} prior to the experiment? We could answer this question in different ways, some of which might appeal to the particular labels that Alice uses to call her outcome. For example, the difference between the largest and smallest possible outcomes that have a non-zero probability is an uncertainty indicator, and it depends on the choice of labels. In principle, Alice could relabel her die's faces to, say, ``a'', ``b'', etc., without changing the essential physical nature of the experiment. We will require our notion of uncertainty to make no distinction between two physically identical experiments that differ only in the outcome labels. In other words, we will consider uncertainty to be a property of just the distribution $\vect p_X$, measured possibly by some real-valued function $\UC(\vect p_X)$. In fact, an even stronger restriction follows. Let $Y$ be a random variable obtained by merely relabeling the different values of $X$. The probability distribution $\vect p_Y$ of $Y$ must then necessarily contain the same values as $\vect p_X$, possibly differing only in their order. Therefore, the effect of any relabeling on $\vect p_X$ is as though the original labels were just permuted amongst themselves: $\vect p_X\mapsto M^{(\pi)}\vect p_X$. In this sense, permutations, although not the only possible way to relabel outcomes, still capture the effect of arbitrary relabelings, as far as our notion of uncertainty is concerned.

If, instead of a die-roll outcome, $X$ were a physical property, e.g. the energy of a quantum harmonic oscillator, arbitrary permutations could result in loss of the variable's physical meaning. To avoid this, we would have to restrict the permutations, e.g. to only shifts in the energy. In general, the restricted class of reorderings is the group $G$ of \emph{symmetries} of the observable underlying $X$, with each symmetry $g$ corresponding to a \emph{change in one's reference frame}. For finite-dimensional observables, $G$ is a subgroup of the group of all permutations.

Our first requirement from a measure $\UC$ of uncertainty is that it be invariant under the symmetry group $G$ of the underlying observable. This immediately leads to the following: Two variables $X$ and $Y$, both representing the same observable, are \emph{equally uncertain} if their distribution vectors are related by some $g\in G$. If $X$ is the outcome of a certain experiment, the random variables $Y$ that are equally uncertain to $X$ include relabeled (under $G$) versions of $X$ (which are perfectly correlated with $X$); the outcomes of other runs of the same experiment with the same apparatus (which may be correlated with $X$ if the apparatus has a memory); outcomes of the same experiment performed on independent but identical apparatuses (uncorrelated with $X$); and in general any $Y$ representing the same observable, with $\vect p_Y=M^{(g)}\vect p_X$ for some $g\in G$.

Thus far, we have found a way to tell when the uncertainties of two variables are equal. Now we will develop a method of determining when and how the uncertainty of one variable can be said to be more, or less, than that of another. To this end, we will first identify certainty-nonincreasing transformations: processes that take any given variable $X$ to an equally- or more-uncertain one, $\tilde X$, by virtue of a ``randomizing'' or ``forgetting'' mechanism. Thereafter, we will use the following rule to compare the uncertainties of two variables $X$ and $Y$ (arbitrary but with the same underlying physical observable): $Y$ is at least as uncertain as $X$ if some uncertainty-increasing transformation of $X$ results in a variable $\tilde X$ that has the same probability distribution as $Y$ up to the symmetries of the underlying observable.

\begin{figure}
  \centering
  \includegraphics[width=0.45\textwidth]{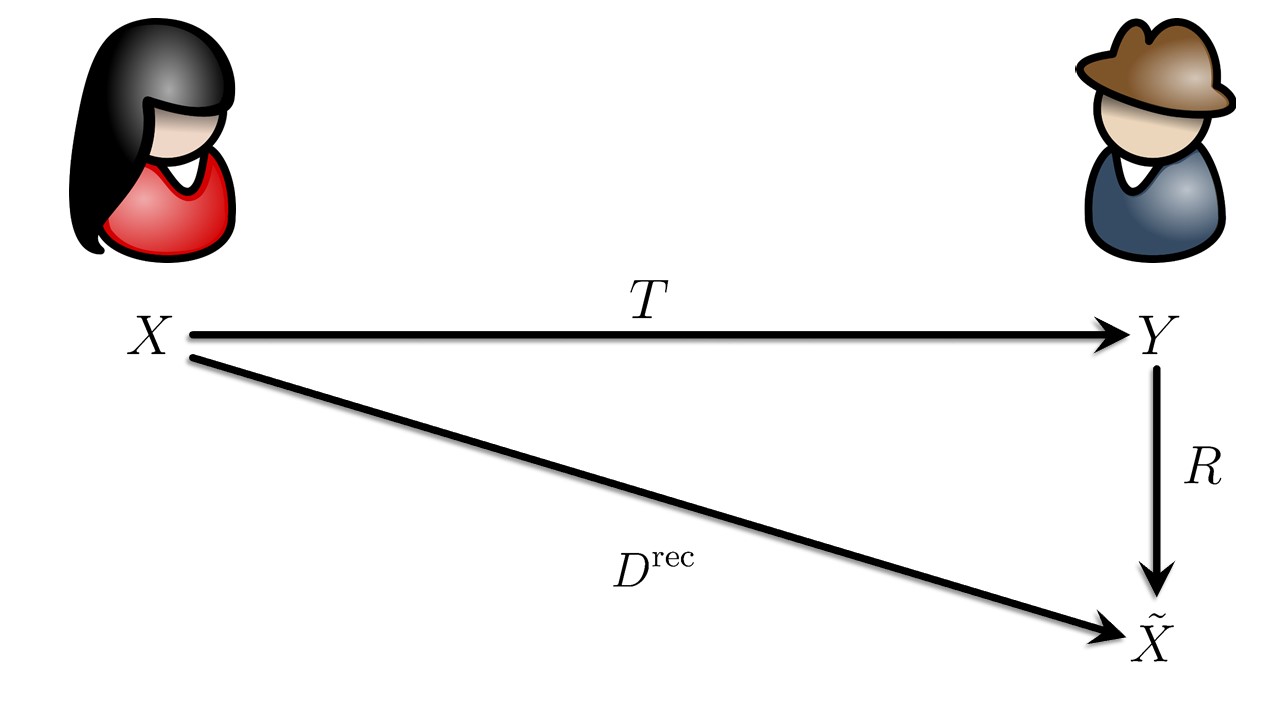}
 \caption{Bob tries to estimate his uncertainty about Alice's variable $X$ after it has been corrupted by channel $T$ into $Y$.}\label{abr}
\end{figure}

In order to identify the certainty-nonincreasing transformations, we will now construct a couple of extended versions of the ``Alice rolls a die'' thought experiment. First, consider the modified experiment depicted in Fig.~\ref{abr} \cite{notecc2}: After rolling her die, Alice sends the outcome $x$ to her collaborator Bob (who doesn't even know the bias distribution of Alice's die) via some classical channel \cite{note4} given by the column-stochastic matrix $T\equiv(T_{y|x})$. Here let's pause to reflect upon the uncertainty in the output $Y$ of the channel. The channel could transmit $x$ perfectly, or with some added noise. In these cases the output $Y$ is equally or more uncertain than $X$. On the other hand, the channel could also completely ignore $x$ and output some constant value, in which case the uncertainty of $Y$ could be \emph{less} than that of $X$. In fact, the processing might result in information in a fundamentally different form from $X$. For example, Alice could just send the parity of her die outcome to Bob, in which case $Y$ doesn't even represent the same underlying observable as $X$. Therefore, we cannot make a general statement about non-increase of certainty under an arbitrary channel.

However, instead of the uncertainty of $Y$ itself, we can consider the following question: How much information does $Y$ contain \emph{about $x$}? Since $Y$ results from processing $X$ with the possible addition of noise or irrelevant information, it cannot tell us more about $x$ than $X$ does. In order to lend mathematical rigor to this statement, we must extract from $Y$ some variable that has the same physical meaning as $X$, so that we can treat them both on an equal footing. Now let's return to Alice and Bob's experiment: Bob, who knows $T$ but not $\vect p_X$, now tries to recover $x$ from the channel output $y$, which is a priori distributed according to $\vect q_Y=T\vect p_X$. Since this game is being designed to analyze uncertainty about $x$, Bob's aim in his recovery task is not to maximize his chances of guessing $x$ correctly, but rather to faithfully account for the uncertainty that $Y$ contains about $x$. Suppose he sees an instance $Y=y$. This could have resulted from a particular $X=x'$ with conditional probability $T_{y|x'}$. Without knowing the prior $\vect p_X$, Bob's rational guess for the likelihood that $X=x$ (among all the possible $x'$) is given by
\be\label{treco}
R_{x|y}=\frac{T_{y|x}}{\sum_{x'}T_{y|x'}}.
\eq
The resulting distribution of Bob's recovered variable (call it $\tilde X$) is given by the composite action of $T$ and $R$ on $\vect p_X$:
\be\label{preco}
\vect p^{\textnormal{rec}}_{\tilde X}=RT\vect p_X=:D^{\textnormal{rec}}\vect p_X.
\eq
Since $Y$ could contain irrelevant information, its uncertainty cannot be interpreted as ``uncertainty about $x$''. On the other hand, $X$ directly represents $x$, while $\tilde X$ results from extracting out of $Y$ precisely all the information it contains about $x$. Therefore, these two variables both represent the same physical observable as $x$, and their uncertainties directly quantify uncertainty about $x$. This equal physical footing also ensures that their uncertainties can be compared under our rules. This comparison tells us that the uncertainty of $\tilde X$ cannot be less than that of $X$. The cumulative transformation that takes $X$ to $\tilde X$ is therefore a \emph{certainty-nonincreasing transformation}. It can be verified easily that for any column-stochastic $T$, with the corresponding $R$ \cite{notert} constructed as in (\ref{treco}), the matrix $D^{\textnormal{rec}}=RT$ is doubly stochastic. The (necessarily degenerative) evolution of the information about some entity (like $x$), when the representation of this information is subjected to \emph{any} classical processing (represented by the action of the channel), is always via such matrices, whose collection we call $\DC^{\textnormal{rec}}$.

\begin{figure}
  \centering
  \includegraphics[width=0.42\textwidth]{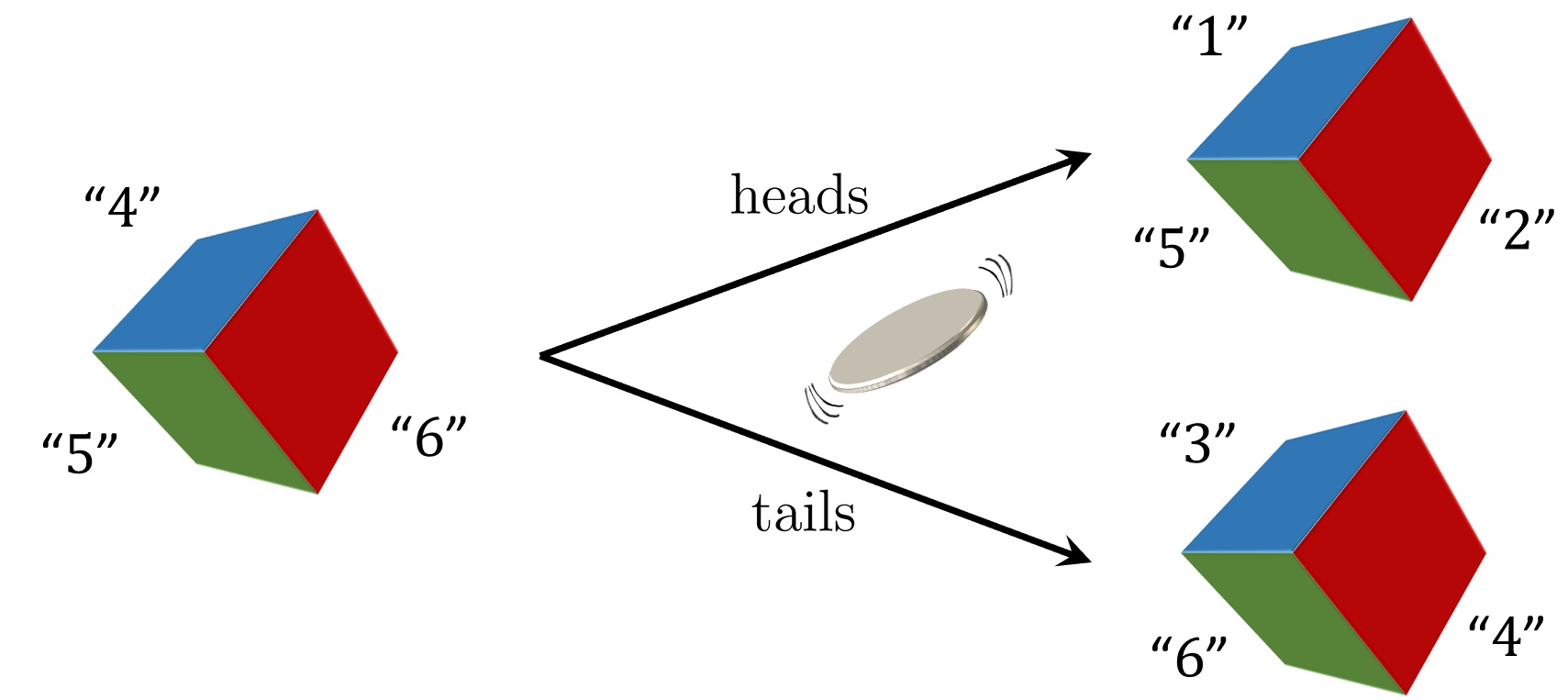}
  \caption{Extended die-roll experiment: Alice relabels her die based on a coin toss, then rolls the relabeled die. The outcome of this experiment is more uncertain than a simple die roll.}\label{ddd}
\end{figure}

We saw that, after the action of a generic channel, the uncertainty of the final variable $Y$ doesn't have a consistent hierarchical relationship with that of the initial variable $X$. In order to draw a consistent rule of certainty non-increase, we had to consider a recovery transformation from $Y$ to $\tilde X$. But there are certain special transformations that always result in certainty non-increase, even without the addition of a recovery transformation. In fact, we already saw an example: symmetry transformations of the underlying physical observable. In the die-roll example, symmetry transformations include non-identity permutations, which can easily be shown to be outside of the die's $\DC^{\textnormal{rec}}$ class, yet result in final variables $Y$ with the same physical meaning and (consistently) no less uncertain than $X$. We will find a family of such certainty-nonincreasing transformations by considering another thought experiment, depicted in Fig.~\ref{ddd}: Before rolling her die, Alice will toss a coin; she will then relabel the die's faces with a permutation that is determined by the outcome of this coin toss, and then roll the die. The random choice of relabeling makes the outcome $Y$ of this modified experiment \emph{more} uncertain than $X$. In general, if a variable $X$ is transformed by applying a $g\in G$ chosen at random under a distribution $\vect t\equiv(t_g)$, the resulting variable $Y$ is distributed as $\vect q_Y=D^{\textnormal{sym}}\vect p_X$, where $D^{\textnormal{sym}}=\sum_{g\in G} t_g M^{(g)}$. Since each $M^{(g)}$ is a permutation, every possible $D^{\textnormal{sym}}$ is doubly stochastic. We denote by $\mathcal{D}^{\textnormal{sym}}$ the set of all such $D^{\textnormal{sym}}$ matrices. If the observable's symmetry group $G$ includes all permutations, then by Birkhoff's theorem \cite{Birkhoff,Bhatia1997}  $\mathcal{D}^{\textnormal{sym}}$ is the set of \emph{all} doubly stochastic matrices, but a restricted $G$ results in a corresponding shrinkage of $\DC^{\textnormal{sym}}$.

The characterization of the classes $\DC^{\textnormal{rec}}$ and $\DC^{\textnormal{sym}}$ is an interesting problem that we leave for future work. While the latter class depends on the symmetry group of the observable, the former depends only on the dimensionality. For a variable with complete permutation symmetry, as noted above, $\DC^{\textnormal{sym}}$ contains all doubly stochastic matrices, in particular all of $\DC^{\textnormal{rec}}$. But under restricted symmetries, each class can contain members not belonging to the other. For instance, take a 3-dimensional variable whose symmetry group is the (order-3) group of cyclic permutations of the components. The two nontrivial permutations are transformations contained (by design) in $\DC^{\textnormal{sym}}$, but not in $\DC^{\textnormal{rec}}$. On the other hand, the matrix
$$\left(\begin{array}{ccc}1&0&0\\0&0.5&0.5\\0&0.5&0.5\end{array}\right)$$
is in  $\DC^{\textnormal{rec}}$, but not in $\DC^{\textnormal{sym}}$. Therefore, the structure of the union of these classes cannot be reduced to either one of the classes. This example can be generalized naturally to higher dimensions.

Due to our restriction to uncertainty comparison between physically-similar variables, the ``sym'' and ``rec'' classes of doubly stochastic matrices together suffice as mechanisms of uncertainty increase. In principle, any function $\UC(\vect p_X)$ meant to measure the uncertainty of $X$ is required to increase under both these matrix classes. But the ``sym'' class is more important that the ``rec'': the former is based on the natural symmetries of an observable, and therefore the constraints that it induces on uncertainty measures are inviolable. On the other hand, ``rec'', even though it is an essential ingredient in the strictest information-theoretic definition of uncertainty, could be ignored in natural situations where information-processing is not involved. Functions that respect the ``sym'' constraints, but violate the ``rec'' ones, nevertheless turn out to be useful indicators of uncertainty. Based on these considerations, we define:\\
\noindent\textbf{Definition 1.} \textit{A \emph{measure of uncertainty} of a variable $X$ is a function $\UC$ of the distribution $\vect p\equiv\vect p_X$ of the variable, satisfying
\be
 \UC(D\vect p)\ge\UC(\vect p)\quad\forall D\in\DC^{\textnormal{sym}};\label{symm}
\eq
\be
\UC(D\vect p)\ge\UC(\vect p)\quad\forall D\in\DC^{\textnormal{rec}}.\label{recc}
\eq
Here the class ``\emph{sym}'' is determined by the symmetries of the variable's underlying physical observable. A function that satisfies (\ref{symm}), but not (\ref{recc}), will be considered a \emph{weak measure of uncertainty}.}

If the symmetry group $G$ of a finite-dimensional $X$ contains all permutations, then functions that satisfy~(\ref{symm}) are called Schur-concave functions \cite{Marshall2011}. Examples of such functions are the entropies of Shannon, R\'enyi, and Tsallis. Now, Hardy \textit{et al.}~\cite{HLP52} proved that the existence of a doubly stochastic $D$ such that $\vect q=D\vect p$ is equivalent to the binary relation $\vect p\succ\vect q$, read ``$\vect p$ majorizes $\vect q$'' \cite{Marshall2011}, which for a general $d$-dimensional vector space is defined as follows. Define $\vect p^\downarrow$ and $\vect q^\downarrow$ as the same vectors with their components arranged in nonincreasing order. Then, $\vect p\succ\vect q$ if, and only if,
\be
\sum_{i=1}^kp^\downarrow_i\ge\sum_{i=1}^kq^\downarrow_i\quad\forall k\in\{1,2\dots,d\}.
\eq
The ``completely certain'' and ``completely uncertain'' distributions $\vect e\equiv(1;0\dots;0)$ and $\vect u\equiv(1/d;1/d\dots;1/d)$ satisfy $\vect e\succ\vect p\succ\vect u$, $\forall \vect p$.

If a variable has restricted symmetries, then the uncertainty hierarchy of its distributions becomes different from the majorization hierarchy. All Schur-concave functions still remain valid uncertainty measures. But in addition, by virtue of the reduction in the class $\DC^{\textnormal{sym}}$, some non{\textendash}Schur-concave functions could also qualify to be \emph{weak} measures of uncertainty [i.e., may violate condition (\ref{recc})]. For example, for a finite-dimensional variable $X$ whose symmetries are cyclic permutations, it can be easily shown that the variance of $X$ is only a weak uncertainty measure. Generalizing the classes $\DC^{\textnormal{sym}}$ and $\DC^{\textnormal{rec}}$ for discrete-infinite and continuous variables may not be straightforward, and we leave it for future work. We expect it to be possible to achieve such a generalization by considering parametrized families of symmetries (e.g. Lorenz transformations) and convex combinations (integrals) over different parameter assignments.

\section{Joint Uncertainty}\label{Sec:Joint}
\begin{figure*}
  \centering
  \includegraphics[width=.95\textwidth]{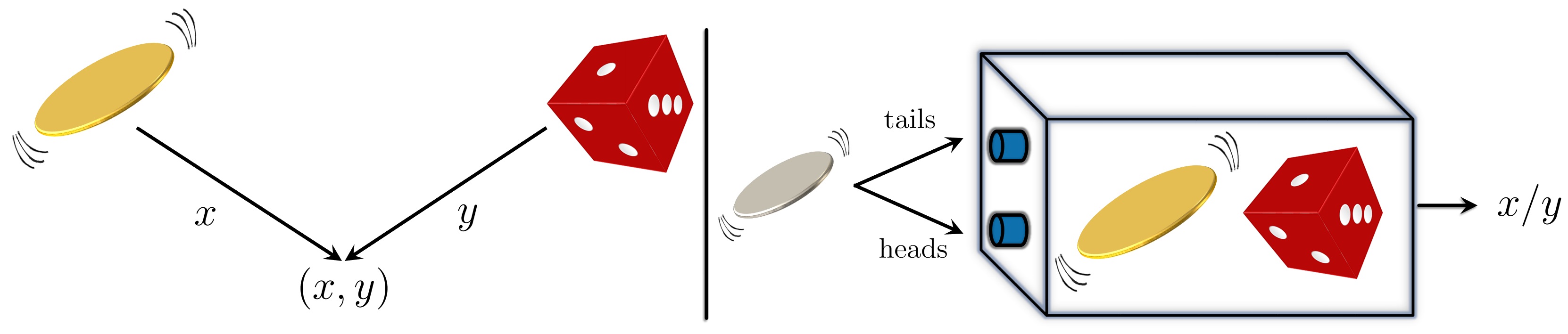}
  \caption{Two possible operational combinations of a coin-toss experiment and a die-roll experiment: (a) Both experiments are performed simultaneously and independently; (b) Only one of the two experiments is performed, based on a random choice.}\label{Fig:TypeAClassic}
\end{figure*}

The uncertainty of the outcomes of \emph{individual} experiments cannot provide a complete description of the quantum uncertainty principle, since most uncertainty relations are lower bounds on measures of the \emph{joint} uncertainty of the outcomes of at least two measurements. For clarity of discussion, here we will restrict to pairs of experiments, each with a finite number of possible outcomes; extension to more experiments is straightforward. To motivate our definition of joint uncertainty, consider the following hypothetical scenarios involving the joint uncertainty of a coin-toss outcome, $X$, and a die-roll outcome, $Y$:

\noindent{\it Example 1}: Perform the combined experiment comprising an independent and simultaneous performance of \emph{both} the original experiments [Fig.~\ref{Fig:TypeAClassic}~(a)]. The outcome is $Z\equiv(X,Y)$, which has $|X||Y|=12$ possible values, distrubuted as $\vect{p}_{Z}=\vect{p}_{X}\otimes\vect{p}_{Y}$. Therefore, $\UC(\vect{p}_{X}\otimes\vect{p}_{Y})$, for $\UC$ any single-variable uncertainty measure (in the sense of Def.~1), serves as a \emph{joint} uncertainty measure of $X$ and $Y$. Most measures considered in the literature on the quantum uncertainty principle, e.g. the sum of Shannon entropies of the individual outcome distributions, can be interpreted through such a combined experiment.

\noindent{\it Example 2}: This time we first toss a second coin to make a choice between the actions ``toss the coin'' (resulting in outcome $X$) and ``roll the die'' (leading to $Y$), and then perform only the chosen action [Fig.~\ref{Fig:TypeAClassic}~(b)]. The outcome $Z$ of this experiment has $|X|+|Y|=8$ possible values, whose uncertainty (modulo the uncertainty in the choice of action) is also a manifestation of the joint uncertainty of $(X,Y)$. In this case, if the choice coin is unbiased, $\vect{p}_Z=\left(\frac{1}{2}\vect{p}_X\oplus\frac{1}{2}\vect{p}_Y\right)$ and therefore we get measures of the form $\UC\left(\frac{1}{2}\vect{p}_X\oplus\frac{1}{2}\vect{p}_Y\right)$. The measures of joint uncertainty proposed in \cite{Oppenheim2010} can be interpreted through such a combined experiment.

As these scenarios illustrate, there could be different ways in which experiments could be combined into one super-experiment, the uncertainty of whose outcomes then reflects an aspect of the joint uncertainty of $(X,Y)$. But the essence of joint uncertainty is not quite captured by any one of these joint experiments. In fact, some joint uncertainty measures, such as the functions $H_\alpha(\vect p_X)+H_\beta(\vect p_Y)$ (where $H_\alpha$ and $H_\beta$ are R\'enyi entropies) \cite{Maassen1988}, and even Heisenberg's $\Delta x\Delta p$, cannot be interpreted as the uncertainty of \emph{any} single combined experiment. The quantum uncertainty principle applies also to cases with several \emph{potential measurements}, each a potential (actual or counterfactual) experiment in its own right.
%

These considerations indicate that the notion of joint uncertainty is not bound to the concept of combined experiments. What, then, are the essential properties of a measure of joint uncertainty? Firstly, the pairs $(X,Y)$ that have the least joint uncertainty are ones where both distributions are completely certain. The most jointly-uncertain pairs, on the other hand, are the ones where both variables are completely uncertain. Furthermore, all the measures of the joint uncertainty of $(X,Y)$ are real-valued functions of the distributions $\vect p\equiv \vect p_X$ and $\vect q\equiv \vect q_Y$, and must reduce to the measures of single-variable uncertainty (as in Def.~1) if one of the vectors $\vect p$ and $\vect q$ is kept fixed. This brings us to the following definition:\\
\noindent\textbf{Definition 2.} \textit{A measure of joint uncertainty of two variables $X$ and $Y$ is a real-valued function $\JC$ of $(\vect p,\vect q)\equiv(\vect p_X,\vect q_Y)$, such that
\be\label{jmono}
\JC\left(D_1\vect p,D_2\vect q\right)\ge\JC\left(\vect p,\vect q\right)
\eq
for all doubly stochastic matrices $D_1,D_2$ in the respective ``\emph{sym}'' and ``\emph{rec}'' classes of both variables. As in the single-variable case, we will call functions satisfying (\ref{jmono}) for the ``\emph{sym}'' class, but not for the ``\emph{rec}'' class, \emph{weak measures of joint uncertainty}.
}

It can be verified that this definition applies to entropic joint uncertainty measures of the form $f(\vect p)+g(\vect p)$, where $f$ and $g$ are single-variable uncertainty measures. The vast majority of the literature on entropic uncertainty relations uses such measures. Note that if the symmetry groups of both variables are the respective full permutation groups, then $D_1$ and $D_2$ can be \emph{any} two doubly stochastic matrices of appropriate dimensions. In this case, the relation in~(\ref{jmono}) states that $\JC$ is monotonic under the \emph{direct product relation} ``$\succt$'' defined by:
$$(\vect p_1,\vect q_1)\succt(\vect p_2,\vect q_2)\Leftrightarrow(\vect p_1\succ \vect p_2\textnormal{ and }\vect q_1\succ \vect q_2).$$

\section{The quantum-mechanical uncertainty principle}\label{Sec:Principle}
The ``uncertainty principle'' of quantum mechanics is actually a collection of identities known as uncertainty relations (UR's), all concerning the uncertainties of individual quantum-mechanical measurements, as well as joint uncertainties of sets of two or more (actual or counterfactual) measurements. Broadly, there are three different operational contexts of UR's: different measurements applied on the same quantum state (either counterfactually or by preparing many copies of the same state); simultaneous (approximate) execution of several measurements; and sequential execution of several measurements. The notions that we developed in the last two sections can be applied in all of these contexts, since they all include instances of finite-dimensional classical variables. But here we will focus on the first type of situation, where different measurements are considered on identical preparations. Furthermore, we restrict to UR's that involve only the probability distributions of measurement outcomes, and not the ``values'' assigned to the outcomes.

Since these UR's involve only the probabilities of outcomes, a positive-operator{\textendash}valued measure (POVM) description of measurements is adequate in the formalism. Consider the case of two POVM's $\AC\equiv\{\Pi_a\}_a$ and $\BC\equiv\{\Gamma_b\}_b$. For a quantum state $\rho$, measurement $\AC$ leads to outcome probability distribution $\vect p(\rho)$ where $p_a(\rho)=\tr{\Pi_a\rho}$, and $\BC$ to $\vect q(\rho)$ with $q_b(\rho)=\tr{\Gamma_b\rho}$. For a so-called \emph{incompatible} pair of POVM's $(\AC,\BC)$, there is no $\rho$ that results in both $\vect p(\rho)$ and $\vect q(\rho)$ completely certain, leading to the existence of a ``minimal joint uncertainty''. Many UR's are statements to this effect:
\be\label{urel}
\JC\left(\vect p(\rho),\vect q(\rho)\right)\ge c\;\;\;\forall\rho,
\eq
where $\JC$ is a measure of joint uncertainty, and $0<c\le C_\JC\left(\AC,\BC\right):=\min_{\rho}\JC\left(\vect p(\rho),\vect q(\rho)\right)$. In some relations (e.g. Robertson's), $c$ is not a constant but rather a non-negative function of $\rho$. The disadvatage of such a lower bound is that it can be zero in some cases even if $\AC$ and $\BC$ are incompatible. For this reason, state-independent $c$'s are favored in most of the recent literature.

In general, our analysis of uncertainty and joint uncertainty enables us to unify the understanding of all UR's of the form
\be\label{uprin}
\JC\left(\vect p_1(\rho),\vect p_2(\rho)\dots,\vect p_n(\rho)\right)\ge c,
\eq
where $\JC$ is a (strong or weak) joint uncertainty measure (under a generalized version of Def.~2) of the $n$ probability distributions $(\vect p_1\dots,\vect p_n)$ that result from measurements $(\AC_1\dots,\AC_n)$ (counterfactually) applied to the same state $\rho$. A vast number of UR's reported in the literature, including most entropic UR's, take this form. In fact, most of the entropic UR's found so far fall under a much stronger restriction. As we mentioned in the previous section, they can all be constructed upon specific notions of joint uncertainty based on the ``combined experiment'' scenarios where either all the measurements are performed on independent, identically prepared quantum systems [as in Fig.~\ref{Fig:TypeAClassic}~(a)], or a random choice is made to decide which of the several measurements to perform [as in Fig.~\ref{Fig:TypeAClassic}~(b)]. All entropic relations based on joint uncertainty measures of the form $f(\vect p)+f(\vect q)$, where $f$ is an entropy function that is additive under tensor products, fall under this category. Going beyond these operational notions and using our general definition of joint uncertainty enables us to construct new UR's, with the following ``recipe'':
\begin{enumerate}
\item Find a measure of the joint uncertainty (under a restricted class of symmetries, if applicable) of the desired number $n$ of distributions, based on Def.~2;
\item For the given $n$ measurements, find a lower bound on the $n$-joint uncertainty of the outcome distributions of the measurements applied to quantum states, like the $c$ in (\ref{urel}). This bound leads to an assertion of the form (\ref{urel}), i.e. an uncertainty relation.
\end{enumerate}

As an illustration, we derive an uncertainty relation for two rank-1 projective measurements on pure states of a 2-level system, using the following joint uncertainty measure constructed using Def.~2: $\JC_2(\vect p,\vect q)=1-\vect p^\downarrow\cdot \vect q^\downarrow$. Here ($\cdot$) denotes the usual dot product. Note that this measure of joint uncertainty is \emph{faithful} in the sense that it is zero if and only if both vectors $\vect p$ and $\vect q$ are completely certain. We find that, for $\vect p$ and $\vect q$ the outcome distributions of the projective measurements with respect to two arbitrary orthonormal bases $\{\ket{x_1},\ket{x_2}\}$ and $\{\ket{y_1},\ket{y_2}\}$,
\begin{equation}
\min_{\ket{\psi}}\JC_2(\vect{p},\vect{q}) =\frac{1}{2}(1-\eta^2),
\end{equation}
where $\eta:=\max_{i,j}|\ip{x_i}{y_j}|$. We provide the proof in the Appendix~\ref{aqub}. Since the measure $\JC_2$ cannot be interpreted based on the two combined-experiment scenarios under which most existing entropic UR's fall, or indeed based on any single-experiment scenario, the above UR is nontrivially different from all previous ones. More generally, for $d$-dimensional $\vect p$ and $\vect q$, any joint uncertainty measure constructed as a Schur-concave function the vector $(p_1^\downarrow q_1^\downarrow,p^\downarrow_2q^\downarrow_2\dots,p^\downarrow_dq^\downarrow_d)$ is a valid joint uncertainty measure of $(\vect p,\vect q)$. So is any Schur-concave function of vectors of dimension $k<d$ constructed with the components $(p^\downarrow_1+q^\downarrow_1,p^\downarrow_2+q^\downarrow_2\dots,p^\downarrow_k+q^\downarrow_k)$. These are just a handful of examples that we contrived for illustration, suggesting that a rich variety of UR's could be obtained by allowing joint uncertainty measures that don't yield themselves to interpetation as the outcome uncertainty of any single experiment.

\section{Universal Uncertainty Relations}\label{Sec:Universality}
We could construct various uncertainty relations using the aforementioned recipe, with the given pair $(\AC,\BC)$ and different measures $\JC$. Every relation is stated in terms of a lower bound like the $c$ of (\ref{urel}), which in turn depends on $\JC$. In general, for a given $\JC$ it might be hard to compute such a bound. But suppose there were a fixed pair $(\vect u,\vect v)$ of distribution vectors, such that
\be\label{unirel}
\JC\left[\vect p(\rho),\vect q(\rho)\right]\ge\JC(\vect u,\vect v)\quad\forall\rho,\JC.
\eq
If there \emph{were} such a pair, then for any given $\JC_0$ we would merely have to compute $\JC_0(\vect u,\vect v)$, immediately yielding a bound. In this sense, finding such a pair would amount to finding a plethora of uncertainty relations; therefore, such a pair can be said to constitute a \emph{universal uncertainty relation} for the pair $(\AC,\BC)$ \cite{Friedland2013,Puchaa2013}.

As it turns out, a nontrivial pair satisfying (\ref{unirel}) never exists for any given $(\AC,\BC)$, because the clause ``$\forall\JC$'' in (\ref{unirel}) includes all single-uncertainty measures of $\vect p$ and $\vect q$ alone, leading necessarily to the trivial choice $(\vect u_0,\vect v_0)$, where $\vect u_0\succ \vect p(\rho)$ and $\vect v_0\succ \vect q (\rho)$ for all $\rho$. Such a $(\vect u_0,\vect v_0)$ would be unhelpful in that it wouldn't impose \emph{joint} restrictions on $(\vect p,\vect q)$. In order to avoid this triviality, we can relax the condition ``$\forall\JC$'', and instead require the inequality in (\ref{unirel}) to only hold for some restricted class of $\JC$'s.

Here we consider again the two restricted combined-experiment scenarios that we discussed in Section~\ref{Sec:Joint}. In the first scenario, \emph{both} $\AC$ and $\BC$ are carried out independently of each other on copies of the \emph{same} state $\rho$. The restricted class of joint uncertainty measures then consists of functions of the form $\JC\left(\vect p(\rho),\vect q(\rho)\right)=\UC\left(\vect p(\rho)\otimes \vect q(\rho)\right)$. In Ref.~\cite{Friedland2013,Puchaa2013}, it was shown that for any given $\AC,\BC$ there exists a distribution vector $\vect u\equiv\vect\omega(\AC,\BC)$ such that the pair $(\vect u,\vect e_2)$ forms a universal uncertainty relation under this restricted class of joint uncertainty measures.
%

Similarly, following Example 2 of Section~\ref{Sec:Joint}, we can consider a combination wherein we first pick, at random, only one of the two measurements $\AC$ and $\BC$, and then perform that one. The joint uncertainty measures considered here are of the form $\UC\left(\vect p(\rho)\oplus\vect q(\rho)\right)$. A nontrivial $(\vect u,\vect v)$ for this restricted class can be found using the methods in \cite{DSum,DSC}.

It might be possible to unify the spirit of the above two classes of universal relations into a larger class, by including all measures of joint uncertainty that are \emph{symmetric} in the two (or more) distributions: $\JC(\vect p,\vect q)=\JC(\vect q,\vect p)$. This requirement avoids the case of trivial relations resulting from the requirement $(\vect u,\vect v)\succt(\vect p,\vect q)$, but we leave it open whether a nontrivial $(\vect u,\vect v)$ can be found. Another way of unifiying several classes of universal relations, each with its respective $(\vect u_i,\vect v_i)$, is by bounding any measure $\JC$ as follows:
\be
\JC\left(\vect p(\rho),\vect q(\rho)\right)\ge\min_{j\in\{1,...,m\}}\JC(\vect u_j,\vect v_j) \quad\forall\rho,\JC.
\eq
An interesting open problem is whether there exists a finite integer $m$ such that minimizing over all $j\leq m$ provides a nontrivial bound for all nontrivial joint uncertainty measures.

Universal uncertainty relations are a powerful tool inasmuch as they generate a variety of uncertainty relations, but the bounds they yield may not be tight. Besides, there are joint uncertainty measures that may not lend themselves to inclusion in a class that admits a nontrivial universal relation, but nevertheless do provide a nontrivial uncertainty relation. An example is the measure $\JC(\vect p,\vect q)=1- \vect p^\downarrow\cdot \vect q^\downarrow$, for which we found a UR in the previous section.

\section{Conclusion}
In this paper, we identified the most basic, measure-independent elements of the concept of uncertainty as applicable to finite-dimensional classical variables. We based our analysis on an information-theoretic study of mechanisms of uncertainty increase: randomly-chosen symmetry transformations; and classical processing via channels (followed by recovery). Corresponding to these, we identified two classes of doubly stochastic matrices, $\DC^{\textnormal{sym}}$ and $\DC^{\textnormal{rec}}$. Uncertainty measures in the strictest sense must be monotonically non-decreasing under both these classes.

We then took a similar information-theoretic approach to the concept of joint uncertainty of several variables, resulting in the principle that the most basic features of joint uncertainty measures must not depend on specific operational combinations of the variables. We then considered quantum uncertainty relations (UR's) of the preparational uncertainty type, where past works have always considered specific operational combinations. Applying our new notion of joint uncertainty not only resulted in a unified understanding of a large class of UR's, but also opened up the possibility of deriving a new class of preparational UR's, namely identities that are mathematically valid for any preparation, but cannot be interpreted based on any single experimental scenario. To illustrate, we constructed a class of joint uncertainty measures with this property, and derived a new UR using one of these measures as an example. Finally, we found that so-called universal uncertainty relations cannot be found over all possible measures of joint uncertainty. We connected universal relations found in past works \cite{Friedland2013,Puchaa2013,DSum,DSC} with specific operational interpretations of joint uncertainty.

In cryptographic tasks we must consider the uncertainty of systems that could be correlated with quantum memories in adversarial control; our recent work \cite{CUP} is a step towards developing a measure-independent notion of such \emph{conditional uncertainty}. More generally, a formalism for treating the uncertainty of \emph{quantum} information correlated with quantum memories is not yet developed. A more complete characterization of uncertainty on infinite-dimensional systems is another challenging future project. This could impact applications of squeezed states, which are ubiquitous in quantum information processing with continuous variables. Yet another open problem is to improve our understanding of universal uncertainty relations; in particular, to answer the open questions posed at the end of Section~\ref{Sec:Universality} regarding stronger classes of universal relations. Finally, there is much to be understood about the classes $\DC^{\textnormal{rec}}$ and $\DC^{\textnormal{sym}}$ of doubly stochastic matrices.

\section{Acknowledgments}
GG is grateful for many interesting discussions with Micha{\l} Horodecki, Amir Kalev, Iman Marvian, and Rob Spekkens. In particular, we are grateful to Iman Marvian and Rob Spekkens, for pointing out to us the role of symmetry in quantum uncertainty relations. The authors acknowledge helpful discussions with Mark Girard, Marco Piani, and Borzu Toloui. We thank Steven Nich for help with the calculations, Patrick Coles and Marco Tomamichel for bringing relevant literature to our notice, and anonymous reviewers for helpful comments and suggestions. This research is supported by the Natural Sciences and Engineering Research Council of Canada (NSERC).

\appendix

\section{Example uncertainty relation for projective measurements on pure qubit states}\label{aqub}
On a two-level quantum system (a qubit) in a pure state $\ket\psi$, consider two rank-1 projective measurements $\AC$ and $\BC$, respectively defined by the orthonormal bases $\{\ket{x_1},\ket{x_2}\}$ and $\{\ket{y_1},\ket{y_2}\}$. When $\AC$ is applied on $\ket\psi$, the resulting outcome distribution $\vect p(\psi)$ has the components $p_1(\psi)=:p(\psi)=\left|\bra{x_1}\psi\rangle\right|^2$ and $p_2(\psi)=1-p(\psi)$; similarly, we denote the distribution of the outcomes of $\BC$ $\vect q(\psi)\equiv(q(\psi),1-q(\psi))$. We shall now find a lower bound on the minimum joint uncertainty of $\left(\vect p(\psi),\vect q(\psi)\right)$, over all pure states $\ket\psi$, under the measure
\be
\JC_2(\vect p,\vect q)=1-\vect p^\downarrow\cdot \vect q^\downarrow.
\eq

Note that $\JC_2$ is a valid measure of joint uncertainty as it satisfies the constraints in Definition~2. We can partition the set of all pure states into two subsets $\SC_1$ and $\SC_2$ given by
\begin{align*}
&\SC_1=\{\ket{\psi}|(p\geqslant0.5,q\geqslant0.5) \textrm{~or~} (p<0.5,q<0.5)\};\\
&\SC_2=\{\ket{\psi}|(p\geqslant0.5,q<0.5) \textrm{~or~} (p<0.5,q\geqslant0.5)\},
\end{align*}
where $p$ and $q$ are understood to be $\psi$-dependent. The function $\JC_2$ can be defined piecewise using this partition as
\begin{equation}\label{Eq:2modesofU}
\JC_2(\psi)=\left\{
      \begin{array}{ll}
        p+q-2pq, & \hbox{$\ket{\psi}\in\SC_1$;} \\
        1-p-q+2pq, & \hbox{$\ket{\psi}\in\SC_2$.}
      \end{array}
    \right.
\end{equation}
Modulo a global phase, $\ket{\psi}$ can be parametrized as
\begin{equation}\label{Eq:optimizingstate}
\ket{\psi}=\cos\alpha\ket{y_1} + \expoi{\varphi}\sin\alpha\ket{y_2},
\end{equation}
with $\alpha\in[0,\pi/2]$ and $\varphi\in[0,2\pi)$. Appropriate global phases can be added to the measurement basis vectors so that
\begin{align*}
q(\psi)&=\cos^2\alpha;\\
p(\psi)&=\left|\cos\alpha\cos\beta+\sin\alpha\sin\beta\expoi{\varphi}\right|^2,
\end{align*}
where $\cos{(\beta)}:=|\ip{x_1}{y_1}|$. The minimization of the function $\JC_2$ using its piecewise definition \eqref{Eq:2modesofU} can be done by separately minimizing over $\SC_1$ and $\SC_2$ and then finding the smaller of the two minima. Let us first consider $\SC_1$. Now, it can be verified that the $\varphi$ dependence of the function is through a term of the form $f(\alpha,\beta)\sin^2(\varphi/2)$, so that the minimization can be carried out first over $\varphi$ alone, and then over all $\alpha$. In the cases where $f(\alpha,\beta)>0$, the minimim over $\phi$ is achieved when $\sin^2(\varphi/2)=1$; if $f(\alpha,\beta)<0$, the minimim occurs when $\sin^2(\varphi/2)=0$. In either case, the minimum over $\varphi$, as a function of $\alpha$, takes the form
\begin{align*}
\min\limits_\varphi\JC_2(\psi)&=:J(\alpha)\\
&=\cos^2\alpha+\cos^2(\beta\pm\alpha)-2\cos^2\alpha\cos^2(\beta\pm\alpha).
\end{align*}
Since $J$ is even in $\alpha$, the subsequent minimization over $\alpha$ leads to the same value regardless of whether the positive or negative sign is used in the $\pm$ above. One can check that the minimum is attained at $\alpha=\beta/2$, yielding
\be
\min_{\ket{\psi}\in\SC_1}\JC_2=\frac{1}{2}\sin^2\beta.
\eq
Using similar arguments, we can determine the minimum over the other partition:
\be
\min_{\ket{\psi}\in\SC_2}\JC_2=\frac{1}{2}\cos^2\beta.
\eq
Note that without loss of generality we can take $0\leqslant\beta\leqslant\pi/2$. Comparing the two local minima, we can express the global minimum succinctly as
\begin{equation}
\min_{\ket{\psi}}\JC_2(\vect{p},\vect{q}) =\frac{1}{2}(1-\eta^2),
\end{equation}
where $\eta:=\max_{i,j}|\ip{x_i}{y_j}|$.


\begin{thebibliography}{10}
\bibitem{note4}
A classical channel, also called a stochastic map, is characterized by a final
  alphabet $\{y_1,y_2\dots\}$, together with a column-stochastic matrix
  $T\equiv (T_{y|x})$. Given an initial value $x_0$, the column $(T_{y|x_0})$
  of $T$ consists of the probabilities with which $x_0$ is mapped to various
  $y$'s.
\bibitem{notert}
Note that $R$ is completely determined by $T$.
\end{thebibliography}
\end{document}